\documentclass[times]{aastex6}
\renewcommand{\dh}{\fontencoding{T1}\selectfont{\symbol{240}}} 
\renewcommand{\DH}{\fontencoding{T1}\selectfont{\symbol{208}}} 

\newcommand{\wtf}{KIC 8462852}

\slugcomment{Resubmission}
\shorttitle{Families of Plausible Solutions for Boyajian's Star}
\shortauthors{Wright \& Sigur\dh sson}

\begin{document}


\title{Families of Plausible Solutions to the Puzzle of Boyajian's Star}


\author{Jason T.\ Wright\altaffilmark{1,2,3}, Steinn
  Sigur{\scriptsize\DH}sson}
\affil{Department of Astronomy \& Astrophysics and \\
Center for Exoplanets and Habitable Worlds\\ 525 Davey Laboratory \\
The Pennsylvania State University \\
University Park, PA, 16802, USA}




\altaffiltext{1}{\url{astrowright@gmail.com}}
\altaffiltext{2}{Visiting Associate Professor, Breakthrough Listen
  Laboratory, Department of Astronomy, University of California,
  Berkeley, CA 94720, USA}
\altaffiltext{3}{PI, NASA Nexus for Exoplanet System Science}

\begin{abstract}

Good explanations for the unusual light curve of Boyajian's Star have
been hard to find.  Recent results by Montet \& Simon lend strength
and plausibility to the conclusion of Schaefer that in addition to
short-term dimmings, the star also experiences large, secular
decreases in brightness on decadal timescales. This, combined with a lack of long-wavelength
excess in the star's spectral energy distribution, strongly constrains scenarios involving
circumstellar material, including hypotheses invoking a spherical
cloud of artifacts.  We show that the timings of the deepest
dimmings appear consistent with being randomly distributed, and that
the star's reddening and narrow sodium absorption is consistent with
the total, long-term dimming observed.  Following
Montet \& Simon's encouragement to generate alternative hypotheses, we
attempt to circumscribe the space of possible explanations with a
range of plausibilities, including: a cloud in the outer solar system,
structure in the ISM, natural and artificial material orbiting 
Boyajian's Star, an intervening object with a large disk, and
variations in Boyajian's Star itself.  We find the ISM and intervening
disk models more plausible than the other natural models.

\end{abstract}





\section{Introduction} \label{sec:intro}

\subsection{Discovery}

\citet{WTF} announced the discovery of an extraordinary star, \wtf, observed
by {\it Kepler} \citep{Kepler} during its prime mission.  First
noticed by citizen scientists as part of the Planet \replaced{Hunter}{Hunters}
project\footnote{\url{http://planethunters.org}}
to examine {\it Kepler}
light curves by eye \citep{PlanetHunters1}, this star exhibited a series of aperiodic
dimming events, with variable timescales on the order of days, amplitudes up
to 22\%, and a complex variety of shapes. \citeauthor{WTF} established that
the data are good and that this behavior is unique to \wtf\ among {\it
  Kepler} stars.

Extensive follow-up by
\citeauthor{WTF} allowed them to determine the star appears to be in all
other ways an ordinary, \added{main-sequence} early F star, showing no signs of IR excess
(that would be indicative of a disk or other close-in material
responsible for absorption) or accretion.  Indeed, the {\it Kepler} field
was above the Galactic Plane, and contains no known star-forming
regions that might produce a star young enough to have significant
circumstellar material.  Using AO, \citet{WTF} did discover a 2\arcsec\
companion consistent with a bound M4V star at a projected distance of
$\sim 900$ au, but this is not unusual, nor does it seem to provide
any explanatory power for the star's {\it Kepler} light curve.

\citeauthor{WTF} constructed ``scenario-independent constraints'' for the
source of occulting material under the assumption that it is
circumstellar, based on the duration and depths of the events, their gradients, the
lack of IR excess, and other considerations.  This allowed them to
rule out many scenarios, and they offered a provisional explanation
that {\it Kepler} had witnessed the passage of a swarm of giant
comets.  

The hypothetical comets, which must be very large to block an appreciable
amount of stellar flux, would only have produced a significant
infrared excess during their periastron passage (presumably around the
time of the {\it Kepler} mission), thus explaining the lack of IR
excess at other times.  \citet{Bodman16} modeled this scenario, and
found it would required hundreds to thousands of comets, perhaps
tidally disrupted from a Ceres-massed progenitor, to explain the final
60 days of the {\it Kepler} light curve.  Despite this success, that
also found that they could not reproduce the long, slow, deep event
observed during {\it Kepler} Quarter 8, casting doubt on the comet hypothesis.

Interest in \wtf\ (which we will refer to as ``Boyajian's
Star''\footnote{The star has picked up other popular monickers,
  including ``WTF'' (ostensibly for ``Where's the Flux?,'' the
  subtitle to the \citet{WTF} paper), and ``Tabby's Star'' (with Dr.\
  Tabetha Boyajian being its namesake). We agree that a more
  memorable name than \wtf\ seems warranted for this extraordinary
  object, and choose ``Boyajian's Star'' in keeping with the long
  astronomical tradition of similar eponyms, such as Barnard's Star,
  Kapteyn's Star, Teegarten's Star, etc.}) increased significantly in
response to popular media accounts of the work of \citet{GHAT4}, who
connected it to the speculation of \citet{Arnold05} that {\it Kepler}
could discover large artificial structures orbiting 
other stars, if they exist. That is, rather than a swarm of comets,
\citeauthor{GHAT4} noted that a swarm of planet- or star-sized structures
would produce numerous transit anomalies, including arbitrary ingress
and egress shapes, anomalous transit bottom shapes, variable depths,
and aperiodicity---all of which characterize Boyajian's Star's
dimming events. \citeauthor{GHAT4} recommended that, until Boyajian's Star's light curve
had a more plausible natural explanation than had been offered to date, SETI\footnote{The search for
  extraterrestrial intelligence, e.g. \citet{Tarter01}} researchers prioritize it in
their searches for communication from extraterrestrial civilizations.

\subsection{Follow-up}

\citet{Marengo15} and \citet{Lisse15} analyzed {\it Spitzer} and {\it
  IRTF} observations, respectively, taken after the {\it Kepler}
observations, and showed that the lack of IR excess noted by 
\citet{WTF} (based on {\it WISE} data \citep{WISE} taken 
before the {\it Kepler} observations) continued to later epochs.  This
ruled out many scenarios involving a cataclysmic, dust-generating
event in a planetary system that occurred between the {\it WISE} and
{\it Kepler} epochs.  \added{\citeauthor{WTF} showed that the
  gradients of the dips were consistent with material on a circular
  orbit at $\sim10$ au, where in equilibrium it would be quite cool and would escape
  detection at these wavelengths.}  

\citet{Thompson16} found no significant millimeter or submillimeter
emission \replaced{putting}{and put} an upper limit of 
7.7$M_\earth$ on the total circumstellar dust mass within 200
au.  \added{This upper limit rules out a very massive debris disks
  orbiting \wtf. \citeauthor{Thompson16} note that only
  $10^{-9} M_\earth$ of dust is required to explain one of the deepest
dimming events seen by {\it Kepler}, and give an upper limit of
$\sim 5 \times 10^{-3} M_\earth$ for dust on elliptical orbits at the
distances favored by the cometary hypothesis.}

On the SETI front, \citet{Abeysekara16} found no evidence of optical
flashes using the VERITAS gamma-ray observatory. \citet{Harp16} and \citet{Schuetz16}
found no evidence of narrowband radio communication or pulsed laser
emission during a simultaneous viewing campaign with the Allen
Telescope Array and the Boquete Optical SETI Observatory\added{, respectively}.  

Only one part of the analysis of \citet{WTF} showing Boyajian's Star
to be an otherwise ordinary F3V star has been called into question:
\citet{Schaefer16} used archival DASCH photographic plate photometry
\citep{Grindlay12,Tang13} to recover 100 years of brightness
measurements for the field from 1890 to 1989. \citeauthor{Schaefer16}'s thorough analysis showed
that Boyajian's Star ``faded at an average rate of $0.164 \pm0.013$
magnitudes per century,'' which he claimed ``is unprecedented for any
F-type main sequence star'' and ``provides the first confirmation that
\wtf\ has anything unusual'' beyond the {\it Kepler} dips.

This claim, at least as extraordinary as the {\it Kepler} light curve
itself, prompted multiple groups to attempt to confirm or refute it.
\citet{Hippke16} and \citet{Lund16} found that the systematic errors in the DASCH
photometry \replaced{to}{do} not permit measurements at the accuracy claimed by
\citeauthor{Schaefer16}. In addition, \citet{Lund16} found several
{\it other} F stars that they claimed {\it do} 
show such long-term variations in brightness.

\citet{Montet16} performed an analysis of the full-frame images from
the {\it Kepler} mission to determine if the secular dimming continued
into the 21\textsuperscript{st} 
century. They found that Boyajian's Star indeed shows irregular,
monotonic fading at an average rate of $\sim 0.7$ mag per century (four times the
\citeauthor{Schaefer16} average) and was $\sim$ 4\% dimmer at the end
of the mission than the beginning. They also show that many of the F
stars that \citeauthor{Lund16} found to have secular photometric
trends are revealed by {\it Kepler} to in fact be shorter-term
variables and that the secular dimming of Boyajian's Star in the {\it 
  Kepler} data is unique among the $> 200$ stars they studied. 

We agree with \citeauthor{Montet16}'s suggestion that the independent
detection of an extraordinary, secular dimming of Boyajian's Star in
the {\it Kepler} data makes \citeauthor{Schaefer16}'s result from the
DASCH photometry more plausible, and with their assessment that such
dimming finds little or no explanation from the comet hypothesis.  

We are thus left with no good explanation for the dimmings of
Boyajian's Star --- neither the complex, short-period events during
the {\it Kepler} epoch, nor the similar amplitude, secular trends seen
in both the DASCH photometry and the {\it Kepler} full-frame
photometry.

\subsection{Plan and Purpose of This Letter}

\citeauthor{Montet16} conclude their paper by stating that they
``strongly encourage further refinements, 
alternative hypotheses, and new data in order to explain the full
suite of observations of this very mysterious object.''  This Letter's
purpose is to be responsive to their encouragement.  In
Section \ref{Periodicity} we examine the periodicities of the deepest dips,
and in Section \ref{Constraints} we examine the constraints on solutions
imposed by the spectral energy distribution (SED) of Boyajian's Star.   In
Sections \ref{Possibilities}--\ref{Intrinsic} we discuss several 
families of solutions with various degrees of plausibilities in light of
these constraints.  \deleted{Because our purpose is to propose general directions
for future research, we decline to develop these families into
sets of detailed models.}

In this work, we shall use the term ``dimming'' in a generic
sense, to refer to the observed changes in the brightness of
Boyajian's Star on all timescales.  We adopt the term ``dips'' from
\citeauthor{WTF} for the days-long events seen in the {\it Kepler}
light curve, and ``secular'' or ``long-term'' dimming for the changes
in brightness noted by \citeauthor{Schaefer16} and \citeauthor{Montet16}.

In our discussion below, \replaced{will always}{we} assume that the long-term dimming identified by
 \citeauthor{Schaefer16} and \citeauthor{Montet16} is real.  The
 combined $\sim V$-band dimming from the two works is 17\% $= 0.20$ mag, which,
 according to the reddening law of \citet{Fitzpatrick99} for
 interstellar dust, would imply the total extinction across
 the spectrum of Boyajian's Star of 15\%.

\section{Periodicities in the Deepest {\it KEPLER} Dips}
\label{Periodicity}
A potential constraint on the location of the cause of the dimming
is periodicities in the patterns of dips in the {\it Kepler}
light curve.  Periods near 1 year might indicate
obscurers in or near the solar system, with dimming modulated by the annual
parallax.  Other periods would presumably correspond with the orbital
motion of material orbiting Boyajian's Star. 

Indeed, the deepest points of the deepest dips (at {\it Kepler} days
793 and 1523) occur 2.000 years apart, a suspiciously precise interval.  It should
be noted, however, that {\it Kepler} is in an Earth-trailing orbit
with an orbital period\footnote{\citet{Mullally16} and \url{
  http://www.nasa.gov/mission_pages/kepler/spacecraft/index-mission.html};
retrieved 2016 July 27} of 372.5 days.  This means that the dips are separated by 1.96
{\it Kepler} years and that the dips' separation's coincidence with an Earth sidereal year
is not important.

Further, taking the six deepest dips (at {\it Kepler} days
261, 793, 1206, 1496, 1523, and 1568), one finds that they all fall
within a narrow range of phases when folded at a period near 24.2 days,
suggesting a close-in orbital period.  \citet{WTF} paid particular
attention to the possibility of a 48.4 days orbital period with events
occurring at both primary and secondary eclipse.  

To check the significance of this period, we constructed a metric of clustering:
\[
M = \left(\sum_i\sin\phi_i\right)^2 + \left(\sum_i\cos\phi_i\right)^2
\]
where $\phi_i = 2\pi t_i/P$ is the phase of dip $i$ occurring at time $t_i$
when folded at a given period $P$. 

Indeed, a search of 2000 periods evenly sampled in frequency between 10 and
700 days reveals that $M(P)$ is maximized with a value of 32.9 at $P=24.2$ days.  We then
repeated this exercise for 10,000 mock sets of six dips with times
randomly drawn from a uniform distribution with the same range as the
{\it Kepler} time series.  The corresponding set of 10,000 $\max(M)$
values has median 30.4, with 16.5\% of all mock sets having a higher value (i.e. more
significant clustering) than the actual {\it Kepler} dips.  The median
period $P$ that maximized $M$ in each of the 10,000 mock sets was 20.8 days.

From this we conclude that the apparent periods found among the deepest dips
are unlikely to be significant, though we acknowledge that more robust
statistical treatments are likely available.

\section{Constraints From Other Data}
\label{Constraints}
\subsection{Optical Constraints}
\label{optical}

The SED and spectra presented by \citeauthor{WTF}, updated with longer-wavelength upper limits by \citeauthor{Thompson16}, put two important
constraints on the obscuring material.

Combining a spectroscopic temperature and metallicity with stellar
models and the observed SED, \citeauthor{WTF} found that Boyajian's Star suffered
$0.11 \pm 0.03$ mag of E(\bv) color excess in 2014, $\sim 1$ year after the
end of the {\it Kepler} data series.\footnote{The
  {\it BVRI} measurements reported in \citet{WTF} were made 2014 June
  6--10 (Kriszti\'an Vida and Tabetha Boyajian, private communication,
  2016)} If this is due to standard interstellar
reddening, this implies $A_V \sim 0.34$, or 37\% $V$-band obscuration
due to dust (corresponding to an optical depth
$\tau_V \sim 1$).

\begin{figure}
\includegraphics[height=3in]{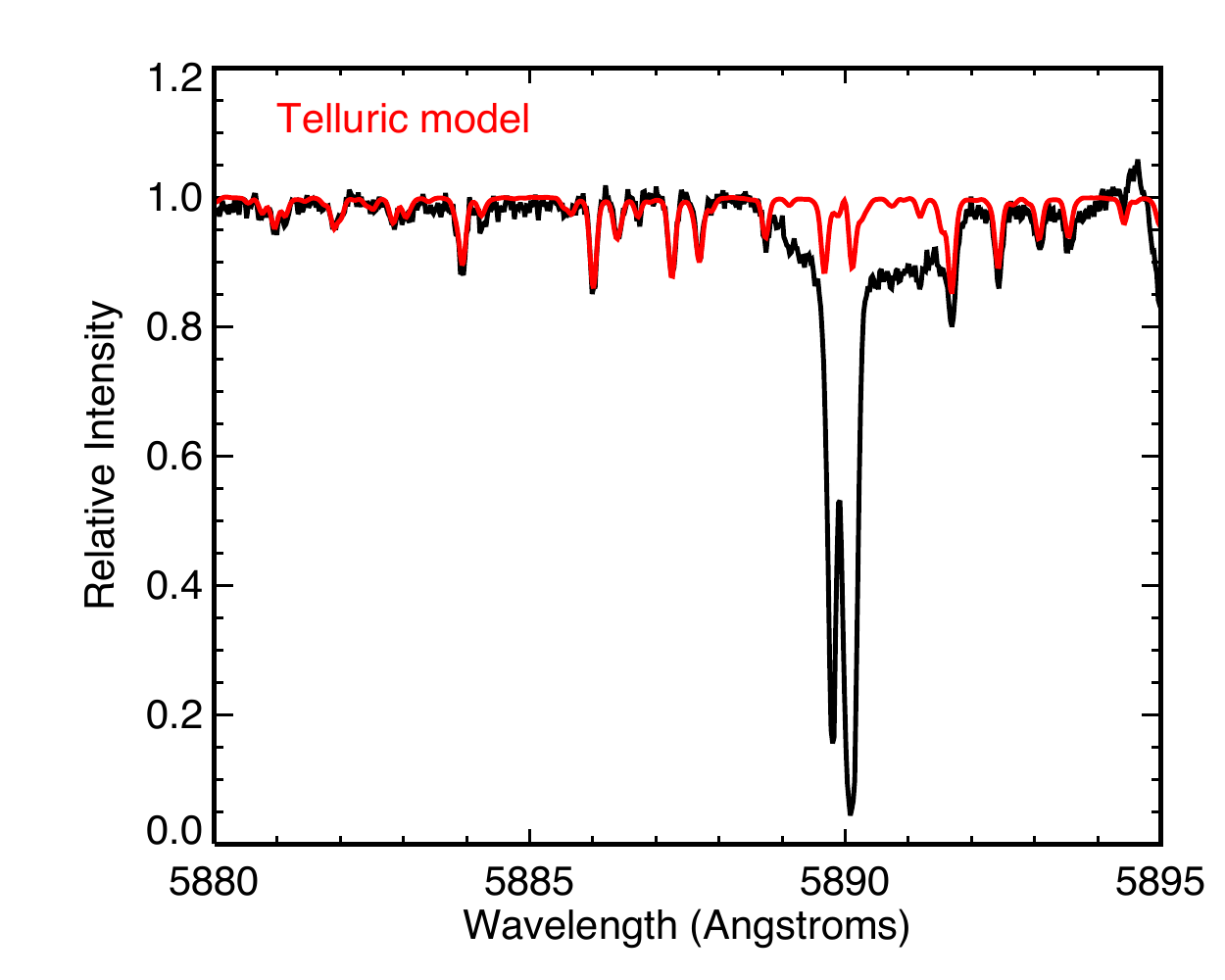}
\includegraphics[height=3in]{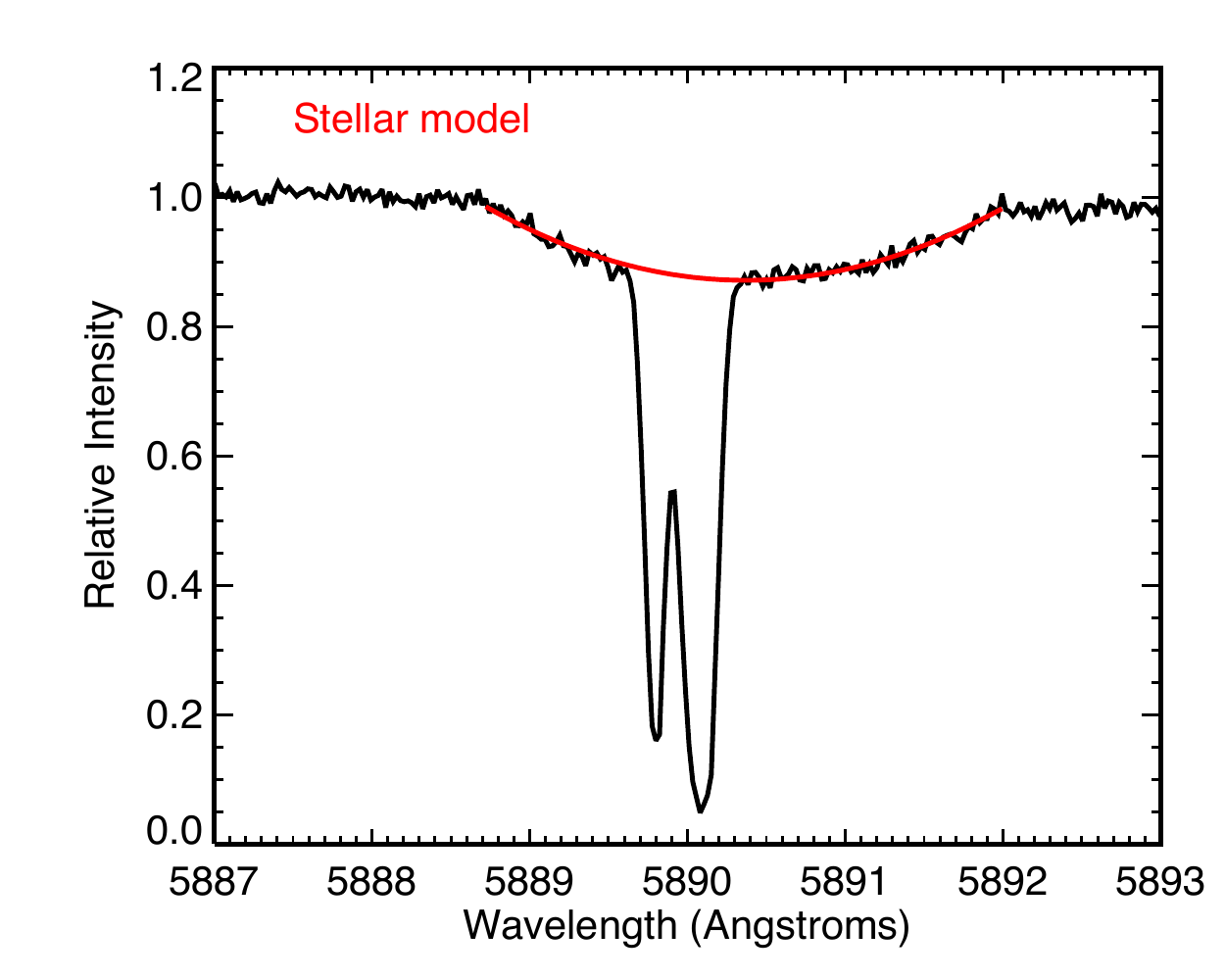}
\includegraphics[height=3in]{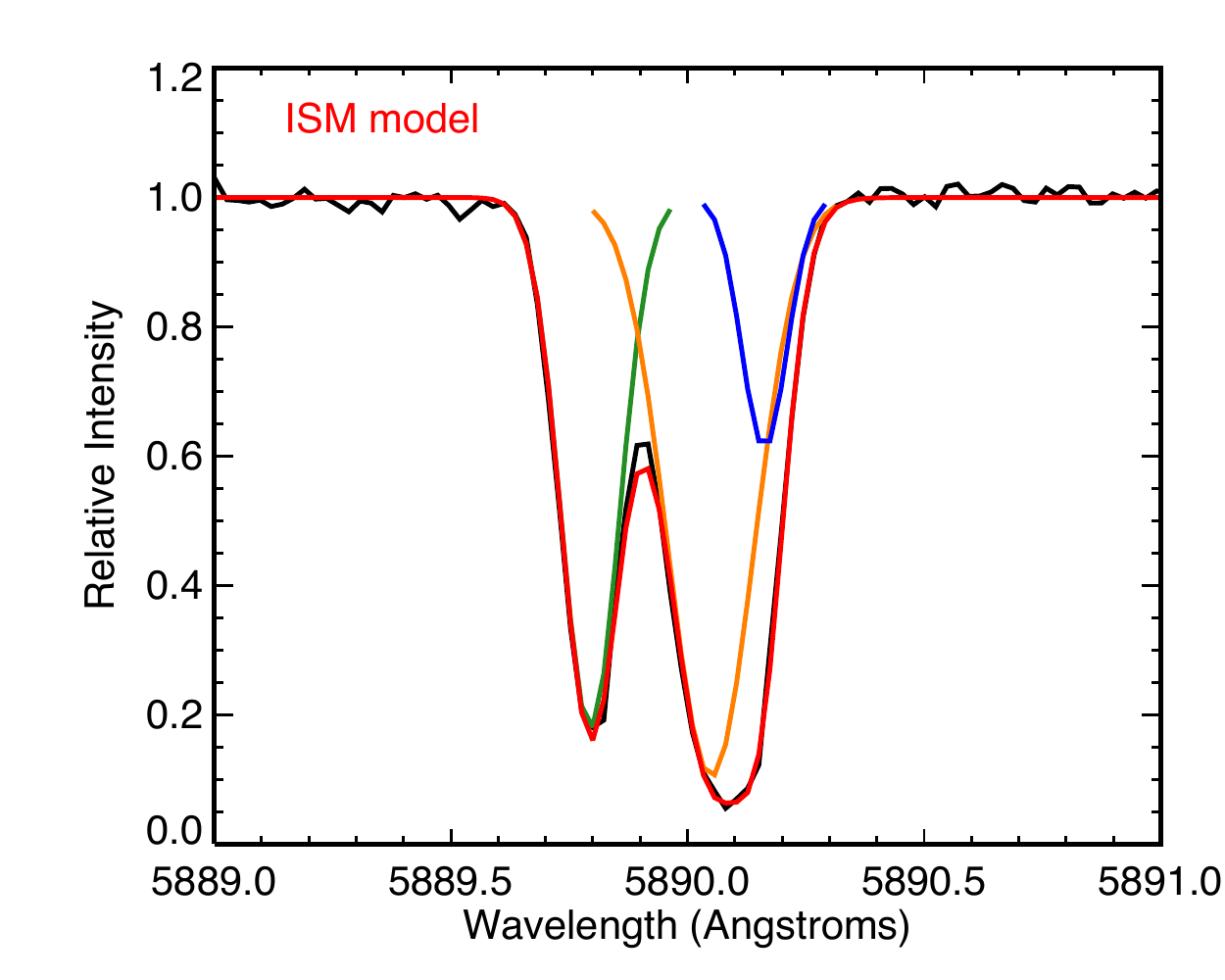}
\caption{Top: Keck/HIRES spectrum of the D2 line of Boyajian's Star,
  with a smoothed \citet{Wallace07} telluric spectrum overlain.
  Middle: A telluric-subtracted spectrum with a simple paraboloidal
  model to the rotationally broadened D2 line.  Bottom: detail of the
  narrow absorption features, modeled with three components.  HIRES
  spectrum courtesy of  B.\
  J.\ Fulton and Andrew Howard.  Plots and analysis courtesy of Jason Curtis.}
\label{D2}
\end{figure}

The narrow sodium absorption features seen by \citeauthor{WTF} in the
spectrum have multiple components, implying multiple clouds of
interstellar material contribute to this extinction.  B.\ J.\ Fulton
and Andrew Howard observed Boyajian's Star with Keck/HIRES
\citep{Vogt94} on UT 2015 October 31.  The D2 line is
well modeled by a smoothed telluric spectrum \citep{Wallace07}, a
parabolic stellar line (which is not physical but is simple and describes the
data well), and three Gaussian components (Fig.~\ref{D2}) having
equivalent widths 170, 200, and 50 m\AA, for a total of 420 m\AA.   

\added{The velocities of the sodium absorption features are offset from the
center of the stellar line by 5--30 km s$^{-1}$.}  If
the secular dimming persisted into 2015, then the lack of a component
at the center of the stellar lines implies that there is not a
persistent, circumstellar absorber with a neutral gas component,
unless it somehow has significant radial motion (so not on a circular
orbit). 

\added{The velocities of the sodium absorption features are offset from each other
by 25 km s$^{-1}$, which exceeds the escape velocity of the Sun at 5
au and is comparable to the escape velocity from Uranus.  They are
thus unlikely to be due to different parts of a single structure and
very likely to be due to interstellar gas.}

It is perilous to extrapolate broadband extinction from sodium absorption
features, which may be unresolved and saturated, and at any rate \replaced{trace}{traces}
only neutral gas, not dust.  But following \added{\citet{Poznanski12},
    who used high- and low-resolution QSO spectra to establish an
    empirical } prescription,\deleted{ of Poznanski et al.\ (2012)} a D2 equivalent
  width of 420m\AA\ corresponds to 
E(\bv)$ = 0.1$, and so is consistent with the observed color excess.

Together, these observations are consistent, and require that
Boyajian's Star be suffering roughly 35\% extinction due to interstellar
dust, and not significantly less or more.

\subsection{IR and Millimeter Constraints}
\label{heat}

The {\it WISE} $W4$ upper limit and the millimeter upper limits by
\citeauthor{Thompson16} are difficult to reconcile with
the amount of long-term dimming seen for scenarios invoking spherical
clouds of circumstellar material.  

\begin{figure}
\plotone{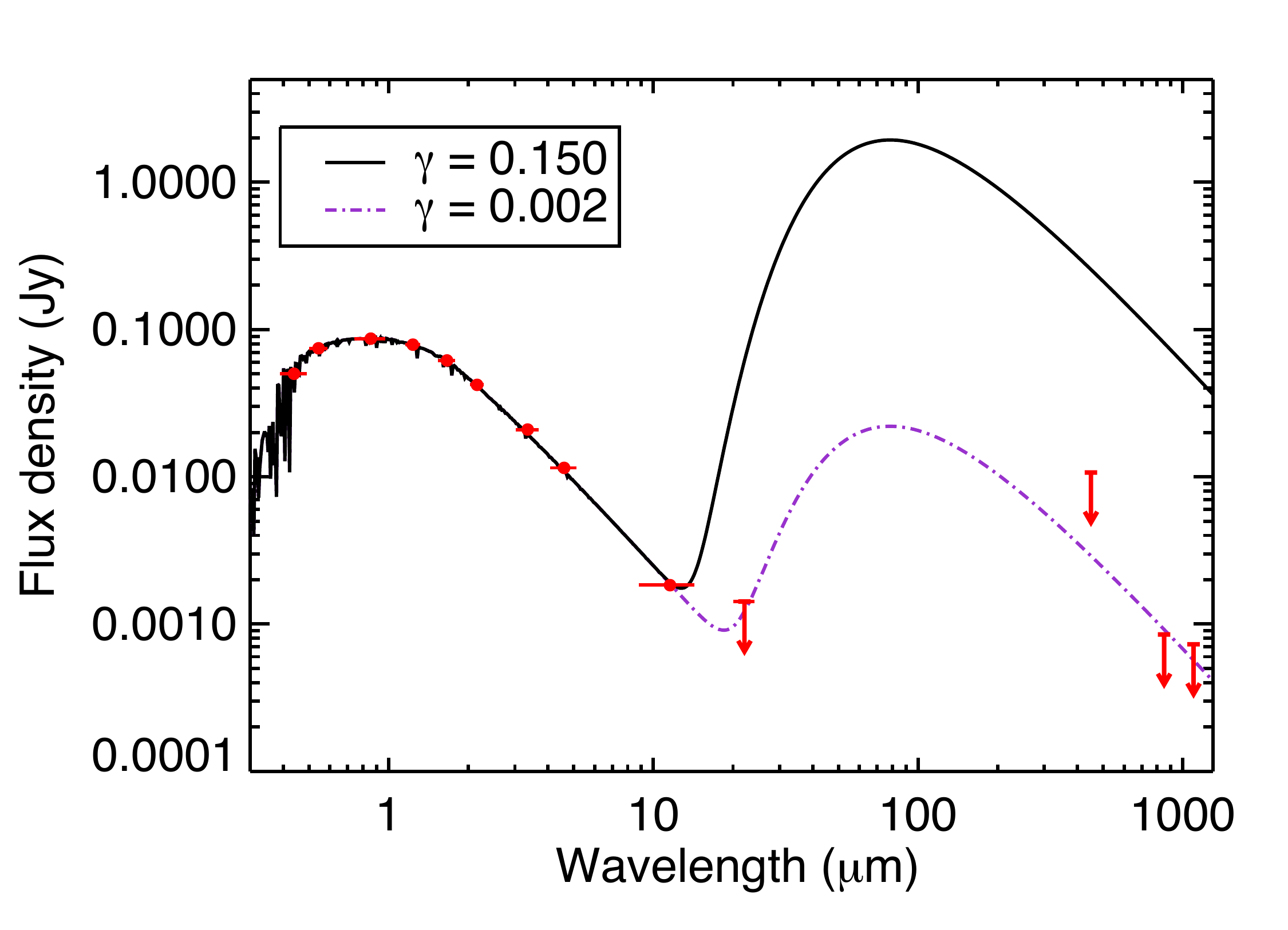}
\caption{NEXTGEN spectrum \citep{NEXTGEN} of Boyajian's Star ($T_{\rm eff}$ 6750 K,
  $\log(g)=4.0$ [cgs], [Fe/H]$=0$), reddened by E(\bv)$=0.11$ following
  \citet{Fitzpatrick99}.  Two cases are shown, with 15\% and 0.2\% of
  the stellar flux
  reprocessed at $T$=65 K (parameterized here
  using the AGENT formalism of \citet{GHAT2} where $\gamma$ is the
  fraction of stellar flux reprocessed).  The measurements and (1$\sigma$) upper limits are taken
  from \citet{WTF} and \citet{Thompson16}.  In both cases, the SEDs are
  scaled to the optical photometry.  The upper limits show that
  less than 0.2\% of the stellar flux is reprocessed as thermal
  emission.  \citeauthor{Thompson16} showed this is equivalent to $< 7.7
  M_\earth$ of dust within 200 au.}
\label{SED}
\end{figure}

The absorbing material must be cold enough not to produce significant 20\micron\ emission, and
have a low enough surface area not to contribute significant 1-mm emission.  
Following the ``AGENT'' parameterization of \citet{GHAT2}, we quantify
the total fraction of stellar flux reradiated at temperature $T$ by
circumstellar material with the parameter  $\gamma$.
As Figure~\ref{SED} shows, the data require $\gamma < 0.2\%$ at $T=65K$.
Other temperatures put even tighter restrictions on $\gamma$ (since all of the longband flux must 
``fit between'' the SCUBA and {\it WISE} upper limits, changing the
temperature in either direction shifts the flux into one of those
bands).

The {\it WISE} and \citeauthor{Thompson16} data thus argue that the
long-term dimming seen by \citeauthor{Schaefer16} and
\citeauthor{Montet16} cannot be isotropic.  That is, if the 
\citeauthor{Schaefer16} plus \citeauthor{Montet16} dimming of $\sim 17\%$ occurs for observers
along all lines of sight, then the absorbing material would easily be
detectable at millimeter wavelengths, regardless of its temperature.  Only circumstellar material
intercepting less than $0.2\%$ of the stellar flux --- for instance, in a
disk --- is consistent with the data.

{\it James Webb Space Telescope} ({\it JWST}) \replaced{photemetry}{photometry} between 20\micron\ and 1000\micron\ would further 
constrain such circumstellar scenarios.

\subsection{Photometry of Nearby Stars}

\begin{figure}
\plotone{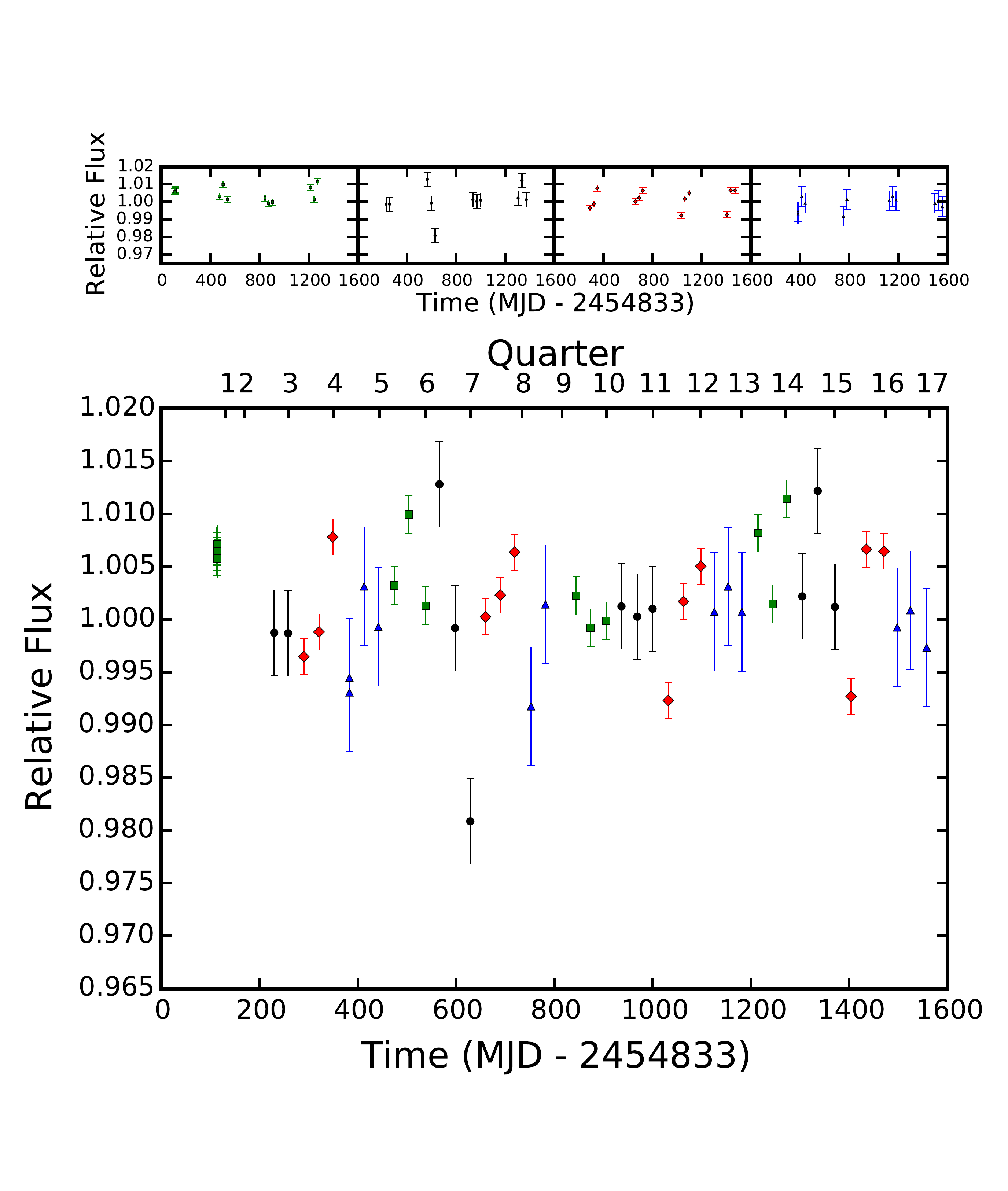}
\caption{Figure kindly provided by Benjamin Montet, showing the
  photometric time series for KIC 8462860, 25\arcsec\ NNW of
  Boyajian's Star, analyzed in the same manner as Boyajian's Star in
  \citet{Montet16}.  This star shows no significant secular dimming, 
  constraining the size of any cloud of material in this direction.}
\label{Montet} 
\end{figure}

Classes of \replaced{solution}{solutions} that invoke solar system or
interstellar material (Sections \ref{SS} and \ref{interstellar}) 
would find support if stars near Boyajian's Star show
similar photometric effects.  \citeauthor{Montet16} examined the brightnesses of
stars in the {\it Kepler} field $\sim $1--5\arcmin\ away.  The lack of
secular dimming or dips in these stars constrain such a
hypothetical cloud to be smallar than this scale. To this end,
Benjamin Montet (private communication, 2016) examined KIC 8462860, a fainter star
25\arcsec\ NNW of Boyajian's star, in the same manner \citeauthor{Montet16} produced
their light curve of Boyajian's star.  Figure~\ref{Montet} shows that
this star shows no significant long-term dimming, showing that if such
a cloud exists, it does not extend this far to the NNW from Boyajian's Star.

\added{Of course, without distance information it is unclear whether KIC
8462860 would be in front of or behind any interstellar absorbers
responsible for the dimming from Boyajian's Star.  Nonetheless,} we
encourage ongoing and future studies of Boyajian's Star to 
similarly monitor neighboring stars for signs of dimming that might
support such hypotheses.

\section{Possibilities}
\label{Possibilities}

The fact that Boyajian's Star is rare --- \citet{LaCourse16} finds no
similar star among the $\sim$ 150,000 {\it
  Kepler} prime mission stars or any of $\sim$ 165,000 {\it K2}
targets  --- suggests that the correct solution is an inherently
unlikely one.  This unlikelihood may help explain why few or no plausible
solutions to the puzzle have yet been offered in the
literature. We will therefore attempt a ``clean-sheet'' survey of
possible solutions, employing a higher threshold for dismissal than
one might normally use.

Occum's Razor points toward a single explanation for both the secular
dimming and the dips.  Of course, in principle they may be unrelated, in which case Boyajian's Star is
extraordinary for two independent reasons.  The possibilities we
explore below will for the most part be general enough that they could
explain either effect, but we will favor explanations that could
plausibly cause both and in each case discuss how this could happen.

There are four possible locations for the dimmings:
in the solar system, in interstellar space, orbiting Boyajian's
Star, and Boyajian's Star itself.  These lead to six broad categories of solutions: instrumental
effects, material orbiting the Sun, an interstellar cloud or disk,
circumstellar dust, other circumstellar material such as
artifacts, and a rare form of stellar variability.  

In the following sections, we discuss families of solutions in each of
these categories. Since our purpose is to circumscribe the space of
plausible solutions, and since Boyajian's Star may be an example of
a new astrophysical phenomenon for which no good precedent exists, we
will usually avoid committing ourselves to a particular model,
preferring to consider families of solutions generally, even at the
cost of being somewhat vague. 

\section{Instrumental effects}
\label{Instrumental}

\citet{WTF}\deleted{and}carefully ruled out instrumental effects in the {\it
  Kepler} camera or data pipeline. The significant public attention to
the star created a renewed interest in the data, and other groups,
such as \citeauthor{Montet16}, have independently reproduced the signal from
the public {\it Kepler} data.  All pixels with significant flux from Boyajian's Star show
similar features, ruling out a problem with any individual pixel.
Dips were observed when the star was being monitored
by four separate modules as the spacecraft rotated, ruling out a
problem with any particular module.  

The long-term dimming in the DASCH data may be more easily explained as
instrumental --- indeed \citeauthor{Hippke16} and \citeauthor{Lund16} argue this
dimming is entirely instrumental --- but the care of
\citeauthor{Schaefer16}'s analysis and the confirmation of secular
dimming by \citeauthor{Montet16} suggest it is real.

\section{Solar system obscuration}
\label{SS}

Material in orbit around the Sun suffers significant
parallactic motion: at a distance of 10,000 au, the annual motion of
the Earth (or {\it Kepler})  corresponds to a parallax of 20\arcsec.
In order to persistently dim Boyajian's star and explain the secular dimming, the material must span
this angle, and so be $\sim$ 1 au across at that distance.  The material's own orbital velocity is
smaller by $\sim 1/\sqrt{a}$ or  $\sim 100$ at that distance,
consistent with the effects being observable for 100 years.  Material
closer than this distance must span a proportionally larger space to
persist on annual and century timescales.  Material much farther than
100,000 au should be considered interstellar (see Section \ref{interstellar}).

If the obscuring material in the solar system, then the variations in the brightness of Boyajian's Star
would be due to a varying optical depth along our line of sight
through the
cloud as the Earth and {\it Kepler} orbit the Sun.  The dips would thus probe
the structure of the cloud on scales of hundredths of an au, while the
secular dimming described variations in its optical depth on au scales.  In this
scenario, the dips in the {\it Kepler} data are not strictly periodic, as might
be expected for parallactic motion, because the cloud itself has
orbital motion, so each {\it Kepler} year probes a slightly different
part of the cloud.  

This scenario finds weak support in the fact that the two deepest dips
occur 1.96 {\it Kepler} years apart, potentially reflecting time for return
of our the line of sight to a dense part of the cloud, with a 
$\sim$2\% correction for the cloud's orbital motion. The absence of
significant dips 0.98 {\it Kepler} year before and after the day 793
dip would then be difficult to explain.

If such clouds were abundant, we would have expected other examples of
Boyajian's Star to have been discovered in the past, so if this
solution is correct they must be rather rare, but there is no obvious
physical reason that such a cloud could not persist if it were cold.
The mass required \added{to }hold such material within a Hill sphere is not large
--- a small asteroid would suffice. The surface brightness in
reflected sunlight would be extremely low at such a distance,
especially if the dust albedo were low, and at any rate obscured in
low-resolution images by Boyajian's Star itself. 

The primary difficulty with this picture is that we have no reason to
expect that there exist $\sim1$ au-sized, $\tau \sim 1$ clouds of
dust in the far reaches of the solar system, especially clouds
containing $10^{-2}$ au, $\tau \sim 1$ clumps.  This difficulty is only
enhanced by the high ecliptic latitude of Boyajian's Star (+62$\arcdeg$).  The origin and
persistence of such clouds are uncertain to say the least, but might
be created by material being gently ejected by a geyser on a geologically active Kuiper Belt
object or Oort cloud member, or a slow collision between comets.  If
the resulting material was sufficiently cold (30 K) then the thermal
velocity of $\sim 100$ m s$^{-1}$ is consistent with the orbital
speeds at these distance.  A 1 au diameter cloud then lasts for
$10^9$ s (i.e.\ centuries).  

This possibility perhaps should be
explored further, especially if neighboring stars are observed to
exhibit similar dimmings.

\section{Obscuration by Interstellar Dust}
\label{interstellar}

\subsection{Obscuration by dense regions of the ISM}
\label{ISM}

The modest Galactic latitude of Boyajian's Star ($b = +6$) allows for
significant extinction along the line of sight.  \citet{Schlafly11}
estimate of the total E(\bv) in this direction is 0.95 mag,
significantly above the measured value of 0.11 mag, indicating that
most of the dust (and structure in \replaced{that figure}{their extinction map}) lies behind the star. 

\citet{Green15} provide a three-dimensional dust map in this
direction, but it is unreliable at the distance of Boyajian's Star
($\sim 450$ pc, their Figure 8 and Web
interface\footnote{\url{http://argonaut.skymap.info}} shows that
results are only meaningful beyond $\sim 500$ pc).  That said, their
``best-fit'' reddening at a distance of 450 pc is E(\bv)=0.09,
consistent with the observed reddening in 2014.


The fact that other stars suffering from significant interstellar
absorption have never been observed to exhibit behavior similar to
Boyajian's Star would suggest that the ISM is not responsible. Nonetheless, we should
explore how such obscuration could explain the observed data.

The space velocity of Boyajian's Star is $\sim 30$ km/s \citep{WTF}.
Our line of sight to the star thus traverses intervening structure at
similar rates, on the order of au/yr.  The dips in the {\it Kepler} data
would thus imply significant sub-au structure in the ISM.

Small neutral structure in the ISM exists.
\citet{Heiles97} describes how observations of 
``tiny-scale atomic structure'' (TSAS) --- exhibited in angular and temporal
variations in hydrogen absorption features in pulsar and quasar spectra --- are best
explained by the alignment of filaments and sheets of cold (15 K)
neutral material along the line of sight.  \citet{Heiles07} summarized
the canonical values for TSASs, including a neutral H{\sc i}
column of $\sim 10^{18}$ cm$^{-2}$ and length scale of 30 au.  Such
TSAS should be overpressurized with respect to the surrounding
ISM, and so evolve dynamically on a timescale of $10^3$ years. 

These values for the dynamical timescale and physical size are broadly
consistent with the secular dimming exhibited by Boyajian's star, but
these canonical column densities are too low to generate the optical 
depths seen in the dimmings of Boyajian's Star by a factor of $\sim
10^2$ \citep[e.g.][who find $A_V = 2.2 \pm 0.09 \times 10^{21}N_{\rm
  H}$ ]{Guver09} and these sizes are too large to provide a
satisfactory explanation for the dips.  

However, sub-au scale structure would be quite difficult to detect without the
cadence and sensitivity afforded by {\it Kepler}, \replaced{and if the phenomenon extends to even smaller scales 
and higher densities then it could be responsible.}{and so would
probably not have been noticed before.  It is not unreasonable to conjecture that the
phenomenon extends to even smaller scales and higher densities, in
which case it should occasionally lead to the behavior exhibited by
Boyajian's star.}  Indeed, \citet{Stanimirovic03} 
argued that such TSAS could be ``a quite rare
phenomenon,'' which may explain why Boyajian's Star is the first to
show its most dramatic effects on broadband optical extinction.

\subsection{An intervening molecular cloud}
\label{Bok}

Alternatively, there might be a chance alignment with a localized
molecular cloud (as opposed to an overdense filament or sheet).

\deleted{For example, Getman et al.
describe the ``mysterious''
high Galactic latitude cloud CG12, which sits 200 pc above the plane
at a distance of 550 pc (about the same distance as Boyajian's Star).
This cloud, spanning tens of arcminutes (and tens of parsecs) on the sky is
obvious across the EM spectrum thanks to illumination by nearby B
stars, embedded protostars, T Tauri stars, and the other usual tracers
of star formation.}

\deleted{Consider, however, if the what the field would look like if there were
no star formation going on.  In this case, the cool gas and dust would
only be obvious by microwave and radio emission, and by its
extinction via star counts.  If the cloud were significantly smaller ---
say, a small Bok globule, spanning only $\sim$ 0.1 pc --- \added{and
  if it had no active star formation} it would likely
not be known at all.}

\replaced{For instance, the}{The} \citet{Clemens88} catalog of small molecular clouds was selected optically based on examination of
the POSS plates, and was sensitive to clouds smaller than 10\arcmin, typically down to
$\sim $1\arcmin.  \citet{Clemens91} found that the mean radius of these
clouds was 0.35 pc.  The clouds in this catalog cluster near the
Galactic plane presumably both because clouds are intrinsically more
common there and because they are easier to identify in silhouette against the large number of
stars there.

A quiescent Bok globule 0.1 pc $\approx$ 20,000 au across and midway
between Earth and Boyajian's Star would have almost certainly escaped detection.  It
would have a radius of 40\arcsec, and examination of the POSS plates
for Boyajian's Star confirms that the star counts are too low in this
region to clearly reveal such a small object, especially if some of the stars
in the image were foreground to it and the globule were not
spherical.  \added{Such high-latitude clouds exist: \citet{Getman08} describe the ``mysterious''
high Galactic latitude cloud CG12, which sits 200 pc above the plane
at a distance of 550 pc (about the same distance as Boyajian's Star).}

In this case, the secular dimming would be naturally explained by the
changing line of sight to Boyajian's Star through the cloud's slowly
varying radial column density profile,\replaced{.  This explanation has
difficulty explaining the observed dips, however, and so for these we must
invoke small-scale knots within the cloud.}{and the dips
would then be explained by small-scale (sub-au) structure within the
cloud.}

\subsection{Implications of the ISM possibilities}

In both of the above scenarios, the space motion of Boyajian's Star and the Sun
(and, presumably, the cloud itself) changes our line of sight through
the cloud, revealing its structure on sub-au scales, a potentially
interesting development for studies of turbulence in molecular
clouds and pressure in the cold ISM.  Indeed,
Boyajian's star would not be unique in probing an intervening
molecular cloud, as many dark
clouds have background stars that could serve this purpose, and the
{\it K2} mission has explored many of them in its survey of
star-forming regions in the ecliptic.  

As with the possibility of a solar system cloud, these scenarios could
be confirmed if the much fainter stars near Boyajian's Star 
could be confirmed to show similar dimming behaviors, or extreme
reddening from darker parts of the cloud.  Indeed, the 2\arcsec\
stellar companion identified by \citet{WTF} as likely to be a bound M4
dwarf with projected separation 900 au might be such a star, since it
has no published color, proper motion, or spectral information. 

The cloud or ISM sheet or filament might also be revealed via its molecular emission, such as
CO lines, with an instrument with sufficient angular resolution, such
as ALMA.  
At the very least,
\replaced{single-beam}{single-dish} observations may reveal an overdensity of atomic or
molecular material in the direction of Boyajian's Star.

\section{The Disk of an Intervening Object}
\label{intervening}

\subsection{Consistency with Existing Observations}

The intervening material may be even more compact than a Bok globule:
it might be a disk.  Given the duration of the long-term
dimming,\deleted{and the lack of long-wavelength excess}explanations
involving a disk must invoke one that is \replaced{very cold and very large
(100's of au)}{$\sim 10^2$ au across}.  The central object must be at
least on the order of a solar mass (to support such a large disk) and 
nonluminous (or it would \deleted{both }appear in the AO imaging of
\citeauthor{WTF}\deleted{ and it would heat the disk, producing long-wavelength
emission; see Section \ref{heat}}).  The disk must be non-accreting, or else
the central object would be detectable. 

\replaced{The only}{One piece of} parameter space left in this
hypothesis is then a chance 
alignment with the disk of a dim stellar remnant, such as a black
hole, quiet neutron star, or old white dwarf.  The pulsar planets
\citep{Wolszczan92,FordPulsar} provide evidence for disks of various
sizes around neutron stars, and \citet{Perna14} provide theoretical
arguments for long-lived black hole fallback disks, so this
possibility warrants further thought.

We note that the Einstein radius of such a black hole between 
Earth and Boyajian's Star is $\sim 4$ mas, or $\sim 10^3$ times
smaller than the angular size of the disk itself, and too small to provide any
reasonable likelihood of generating a microlensing event.

\subsection{Number Density Required}

We can estimate the number density of such objects in the Galaxy
required for such an alignment to have a reasonable chance of one
detection among 100,000 stars.  Given
that the \replaced{timescales}{timescale} of the obscuration is on the
order of or longer than the
life of the {\it Kepler} mission, the volume of the Galaxy probed by a
single star in the {\it Kepler} field is on the order of
\[ 
V \approx D^3 (r/d)^2 
\]
\noindent where $D$ is the distance to the star, $r \sim $ (30 km/s
$\times 100$ years) $\sim 600$ au is the
physical size of the intervening object, and $d \sim D/2$ is its
distance.  If we estimate that
$N =$100,000 stars at typical distances of $D\sim 500$ pc were tested for similar sorts of
dips, then we have 1 such detection in a search volume $NV$, so

\[
n \sim (NV)^{-1} \sim 10^{-3} {\rm pc}^{-3}
\]

This is a rather reasonable number density, being an order of magnitude
smaller than the number stars at the Galactic midplane \citep[$\sim
10^{-2}$ pc$^{-3}$, e.g.][]{Holmberg00} and perhaps not inconsistent
with the estimated $\sim 10^9$ black holes and 
neutron stars in the Milky Way \citep{Timmes96,Santana16}. This scenario thus
deserves further study, especially if it is found that these sorts of
cold, large disks are quite common and long-lived around such remnants,
or if stellar remnants are more common than the \citeauthor{Timmes96}
estimate. 

\subsection{Plausibility of Black Hole Disks}

Following \citet{Perna14}, we consider a fallback disk around a
canonical 10 $M_\sun$ black hole in the field.  \replaced{The argument scales
for a compact object.}{For a compact object,
the timescales go as $M^{\frac{1}{2}}$.}   The initial fallback consists of a large
amount of mass, 1 $M_\sun$ or more for a black hole, which falls
back with some finite angular momentum and stalls at the
circularization radius, generally on the order of $10^2 R_\sun$. 

The initial fallback material accretes very rapidly onto the central
object, powering a bright X-ray source.  This bright X-ray phase ends
after $\lesssim 10^7$ yr as the inner disk is depleted and internal
angular momentum evolution drives a small amount of material outward
from $100 R_\sun$ to $\sim 100$ au.  The outer disk
is cold, evolving slowly on viscous time scales and internal secular
evolution time scales.    

The viscous time scale, $t_0 \sim 10^3 \alpha_{0.1}^{-1} M_{10}^{-1/2} R_{100}^{3/2}  (R/H)^2 $ yr,
where $\alpha_{0.1}$ is the Shakura-Sunyaev viscosity parameter (written in units of $\alpha_{0.1} = \alpha/0.1$),
$M_{10} = M/(10 M_\sun)$ is the mass of the central object, 
$R_{100} = R/100$ au is the radial distance, and $H$ is the disk scale
height, assumed for simplicity to be constant and $\sim R/100$.

The remnant disk mass is $\lesssim 10^{-2} M_\sun$.
The disk is Toomre stable for masses less than about $10^{-3}
M_\sun$. For disk masses above 10 $M_\Earth$, for solar
metallicities, the disk opacity is of order unity at 100 au. 

At 100 au, the viscous time \replaced{scales}{scale} exceeds $10^8$ years, and the disk
evolves slowly, expanding toward $10^3$ au over the lifetime of the
object, the (metal rich) fallback material will have created dust
that will presumably evolve similarly to the outer solar system disk,
but slowly. Planetesimal growth and secular disk instabilities will
lead to au-scale substructure in the disk, consistent in principle
with the observed dips.  Slower variation in surface density with radial distance
would then explain the secular dimming.

Since the central illumination is near zero by hypothesis, and since
the viscous timescale is long, the heating of the disk is minimal and
it will quickly cool to very low temperatures, well below the upper
limit set by \citeauthor{Thompson16}

Since some fraction of black holes formed by supernovae should have
such fallback disks and since angular momentum conservation demands
an extended remnant disk, the primary remaining question with this
suggestion is their overall frequency and the detailed probability of
having seen one occult a star in the {\it Kepler} field.  

\subsection{Intervening Disk of a Binary Companion}
\label{OrbitingDisk}

A disk similar to the sort discussed in above might
also be found in orbit around Boyajian's Star, instead of in the field
\citep{Cameron71}.  It could then be
smaller than in the interstellar case, having a minimum physical size
$r \sim \tau / v$ set by the  $\tau \sim$ 100 year duration of the \citeauthor{Schaefer16} dimming and its
orbital velocity $v$.  However, in this, case the angular extent leads to a very unlikely
alignment.  If a fraction $f$ stars have companions of this sort,
spanning angular size $r/a$ as seen from the star, then the probability of
an alignment (obscuration) is
\[
p \sim f (r/a)^2/(4\pi)
\]
However by (Johannes) Kepler's Laws, $v^2 \sim 1/a$, and from (the
observatory) {\it Kepler} we have $p \sim 10^{-5}$, yielding 
\[
f \sim 10^{-8} a^3
\]
\noindent with $a$ in units of au.  For wide companions at $a \sim$
$10^3$ au, this yields $f \sim 10$, meaning that even if every star had
a companion capable of producing the dips in Boyajian's Star, we would
only have a 10\% chance of having seen it in the entire {\it Kepler}
data set. 

Decreasing $a$ to 100 au brings this down to $f \sim 1\%$, which is
better but still very implausible and begins to conflict with the
constraints from the lack of infrared excess, and the considerations
presented in \citet{WTF}. 

\added{While these probabilities seem too low to make this scenario
plausible, objects such as EE Cep \citep{EECep,EECep2} and $\epsilon$
Aur \citep{epsilonAur} exist, so it perhaps should not be entirely dismissed. }

Another possibility is that the disk material is associated with the
2\arcsec\ companion identified by \citeauthor{WTF} This appears to be
a found M4 dwarf at projected separation 900au, and so would need to
have a disk at least 900au in radius to be responsible for the dimming
in this scenario, and yet contributes no long-wavelength excess to the
SED of Boyajian's Star, which seems unrealistic.  Boyajian's Star (and
so too, presumably, the M4 companion) shows no signs of youth or
accretion, so an optically thick disk would not be expected around either.

\section{Circumstellar Material}
\label{Circumstellar}

\citet{WTF} discussed circumstellar scenarios extensively and concluded
that material such as comets on eccentric orbits 
could be responsible for the short-term dips.  Before the long-term
dimming of Boyajian's Star was discovered, other possibilities 
invoking ephemeral close-in material were possible, for instance,
material ejected by a close-in planet, similar to KIC 12557548
\citep{Rappaport12}.  

We will not repeat the discussion of \citeauthor{WTF} here, except to
note that explanations invoking any stellar-mass close-in orbital companions
(such as stars or stellar remnants) are ruled out by the lack of radial
velocity variation seen by \citet{WTF}.  Atomic circumstellar material at small orbital
velocity being responsible for the secular dimming is also
inconsistent with the lack of sodium absorption at zero velocity (Section \ref{optical}).

While the asymmetric light curves of some of the dips are reminiscent of the transit profiles of
gravity darkened stars, their depth and lack of periodicity preclude
such transits, and the star's rotation period \citep[which][ measured
from the photometry and spectral broadening]{WTF} precludes
significant gravity darkening.

It is worthy of note that light curves with some qualitative
similarities to Boyajian's Star have been observed for the central
stars of planetary nebulae \citep[e.g.][]{Mendez82,Miszalski11}.
These events are thought to involve the formation of dust in the
planetary nebula and modulated by the binary orbital motion of the central
star.  The lack of any signs of a planetary nebula or similar source
of dust toward Boyajian's Star (such as significant {\it W4} emission
from {\it WISE}) would seem to rule out a similar
explanation for it.  This explanation does bear qualitative
similarity, however, with the interstellar dust explanations proposed
in Section \ref{interstellar}.  Similarly, a class of young stars with
disks known as ``dippers'' have asymmetric and variable-depth dips,
similar to those
seen in Boyajian's star \replaced{(Ansdell et
  al. (2015)}{\citep{Ansdell15,Scaringi16}}.  However, \replaced{unlike
  dippers}{whereas dippers are observed to have strong IR excesses
  from their disks}, Boyajian's Star shows no IR excess belying
close-in dust, and so must have a different explanation. 

The possibility of a cloud in the outer solar system from Section \ref{SS}
has a counterpart in this section in the form of a large cloud with large orbital
period orbiting Boyajian's Star.  The constraints on its size then
come only from its orbital motion, with the effects of parallax being
negligible.  

\section{Artificial Structures}
\label{Megastructures}

\subsection{Fleshing Out the Hypothesis}

Because we have no way beyond fundamental physics to constrain or parameterize the
likelihood or nature of artificial structures that may be orbiting Boyajian's
Star, it is a sufficiently flexible model to fit almost any data.  As such, it
should be an ``explanation of last resort''
\citep{GHAT4}. Nonetheless, until the mystery is solved it is
worthwhile to at least outline \replaced{what that explanation is.}{a
  straw-man version of the hypothesis}.

\citet{Lintott16} suggested, apparently tongue-in-cheek, that the
secular dimming
of Boyajian's Star might be representative of the pace of
construction of a ``Dyson sphere'' orbiting Boyajian's Star
\citep{Dyson60}. A similar possibility is skeptically mentioned by \citet{Villarroel16} with respect to the object
USNO-B1.0 1084-0241525, which seems to have disappeared in the past
few decades.  A simpler hypothesis than such rapid construction rates
exists, however.

For the sake of concreteness, assume the artificial structures to be
thin panels used for stellar energy collection to be used on-the-spot to power
``factories'' performing some task \citep{GHAT1}. The panels might have a range of sizes,
from \replaced{meters (analogous to the panels on our satellites
and interplanetary probes) to of order a stellar radius}{$\sim 1 {\rm m} -- 1 R_\sun$}.  Their
shapes, too, might be arbitrary, or they may orbit in formations.
Finally, they might span a range of orbital distances, and so have a
range of orbital velocities.  They would
thus form a ``swarm,'' with the smallest panels effectively acting as an opaque
screen, larger panels causing the star to flicker as they transit, and
the largest ones causing the dips seen in the {\it Kepler} data. \citep{GHAT4}

If the circumstellar volume is filled with such panels, close-in panels
would shadow panels farther out, reducing their efficiency.  An
optimization balancing total energy collection against total mass or
construction costs might thus result in a typical optical depth of order unity.  The optical
depth along our line of sight would be modulated by the orbital
motions and clustering of the panels. Indeed, if a very large
opaque structure orbited into view the star might entirely disappear \deleted{in the
optical }while it occulted the star \citep{GHAT4}.

The timescale of the variations thus reflect the size scale and
orbital velocity of the absorbers: the years- or decades-long dimming
seen by \citeauthor{Schaefer16} and \citeauthor{Montet16} would be due to
au-scale overdensities of panels in the swarm orbiting into view.  The
days-long events noted by \citeauthor{WTF} would be due to star-sized individual
objects (or tight formations of smaller objects) transiting the disk.  

\subsection{Tests of the Hypothesis}

This\deleted{ particular} hypothesis might find support in at least three ways.

First, as \citet{GHAT4} noted, the panels might be expected to be geometric
absorbers, and so produce achromatic dimmings.  This could be checked
once the total extinction of the star (perhaps as established by stellar models and a GAIA
parallax) can be compared to that expected from the observed color
excess.  If the GAIA parallax is significantly larger than the
spectroscopic parallax \replaced{(that is, if the star is much closer 
than expected because it is being extinguished by much more than the
35\% implied by the color excess)}{after accounting for reddening},
this would imply that geometric absorbers, not dust, are responsible
for a significant fraction of the absorption. 

Conversely, if GAIA finds that Boyajian's Star's brightness is
consistent with its distance and reddening, this implies that the
secular dimming observed by \citeauthor{Schaefer16} and
\citeauthor{Montet16} is entirely due to dust.  Increased reddening
during a future dip with be a further blow against the megastructure hypothesis.

Second, if there is a spectrum of sizes of panels ($f(r)$) and orbital
velocities ($f(v)$), the star should ``flicker'' at timescales
corresponding to $r/v$ and $R_*/v$ with higher amplitudes than typical
F stars.

Finally, of course, communications SETI efforts could confirm the
existence of an extraterrestrial civilization in the direction of
Boyajian's Star, which would strongly support this interpretation of the data.

\subsection{Waste Heat Constraints}

The {\it WISE} and \citet{Thompson16} constraints on long-wavelength
excess of Boyajian's Star put constraints on the collecting area,
temperature, and energy efficiency of the factories.  In the AGENT
parameterization of \citet{GHAT2}, these are $\alpha$ (the fraction of
stellar luminosity absorbed), $T$ (the typical waste heat disposal
temperature, characteristic of the operating temperature of the
factories), and $\gamma$ (the fraction of stellar luminosity radiated
as waste heat).

\citet{GHAT2} argued that values of $T$ much below 150 K
would be surprising, since there is little thermodynamic efficiency to
be gained by going to such low temperatures, but a huge increase in the amount of
collecting/radiating area necessary to collect/dispose of energy at
those temperatures.  \citeauthor{GHAT2} further noted that most work
done by humanity results in $\alpha=\gamma$; that is, little of the
energy we generate or collect ends up stored or emitted as low-entropy
radiation, and virtually all of it is reradiated as waste heat after it
is used.  Of course, neither of these observations are physical limits
on an advanced civilization, but it is interesting to see whether they
are consistent with the data for Boyajian's Star.

In Section \ref{heat} we saw that the  {\it WISE} and
\citeauthor{Thompson16} upper limits implied $\gamma < 0.2\%$. This
is inconsistent with a spherical swarm of collectors being
responsible for the observed secular dimming (which would imply
$\alpha \gtrsim \replaced{17}{15}\%$) unless $\alpha \gg \gamma$.  If we assume that some sort of non-dissipative work is
being performed with the energy, such as the emission of low-entropy
emission (lasers or radio beacons, for instance) or energy-to-mass
conversion, the efficiency of this work is limited by the thermodynamic (Carnot)
efficiency $\eta$ set by the factories' radiation temperature.  For $T = 65K$
and $T_*=6750$ this is $\eta \sim 99$\%, and so if the factories operate at this
limit,  we have 
\[
\gamma = (1-\eta) \alpha \sim 0.15\%
\]
\noindent which is just barely consistent with observations.  That is,
a spherical swarm of megastructures can produce the
\citeauthor{Schaefer16} dimming only if they emit around
$T=65$K and operate near the maximum efficiency 
allowed by thermodynamics. More sensitive measurements between
20--$10^3$\micron\ would rule out this hypotheses completely.

The data are still consistent, however, with structures whose
collection or re-radiation strongly anisotropic: either we are seeing 
obscuration from a ring-like structure of collectors (allowing
$\alpha=\gamma$ to be 100 times smaller) or they preferentially
re-radiate away from our line of sight (which seems unlikely unless
paired with a ring explanation, where the plane of the ring
establishes a preferred radiation direction away
from Earth).

\section{Intrinsic variations}
\label{Intrinsic}
\subsection{Timescales}
\label{Timescales}

Except for stars undergoing pulsations significantly large enough to alter
their internal structure to the point that their core pressure
changes, main-sequence stars' interior luminosities are constant on
timescales shorter than the nuclear timescale.  Their surface luminosities can vary by small
amounts about this constant value --- the Sun's luminosity changes by $\sim 0.1\%$ throughout
the solar cycle --- but any decrease in surface luminosity must
eventually be balanced by a later increase.  The timescales for these
variations are a sound-crossing time (on the order of minutes, as with asteroseismic variations),
that of a driving mechanism (in the case of the solar cycle or pulsating
stars), or a thermal timescale (i.e.\ a Kelvin-Helmholtz timescale,
$\sim 10^6$ yr for an early F star.)  The presence of dimmings in
Boyajian's Star on timescales from days to decades is hard to
reconcile with any of these mechanisms.

\citeauthor{Montet16}'s demonstration that Boyajian's Star has significant secular
dimming on decadal timescales --- and the corollary that
\citeauthor{Schaefer16}'s century-long dimming is therefore likely to
be real --- would therefore seem to rule out any explanations that
involve the star itself.  Further, their demonstration that the {\it
  Kepler} light curve shows only short-term {\it dimmings}, and that
previously reported {\it brightenings} were artifacts of data
processing removing low-frequency power, argues against the source of
the dimmings being changes in the surface luminosity of the star
itself.

\subsection{Polar Spots}
\label{Spots}

One possible way out is to invoke surface inhomogeneities: very large
starspots could create temporary dimmings balanced not by later
brightenings, but by bright regions elsewhere on the stellar disk.
This explanation finds difficulty in the clear, 0.88 day rotation period
of the star \citep[from both the photometry and consistency with the
observed line broadening][]{WTF}; which is much shorter than most of the dips' 
durations.  

Slow growth of large polar spots, as suggested by \citet{Montet16},
might explain the long-term dimming (which would be seen as a
brightening from an edge-on orientation), but would --- as
\citeauthor{Montet16} point out --- be an unexpected and extraordinary
feature for an early F star.  This explanation also is not obviously
consistent with the multiple timescales for the dimmings.

\subsection{A Post-brightening Return to Normal?}
\label{Return}

An alternative to the star being {\it dimmer} than it should be is that it is actually
too {\it bright}, and is returning to an equilibrium state after some
event injected significant energy into its envelope or temporarily
increased its core luminosity.  For instance, perhaps Boyajian's Star
recently merged with a brown dwarf or another star and is still 
processing the absorbed orbital energy.  Residual material from the
merger might transit the star occasionally, producing the dips, or the
star might still be undergoing internal changes on a hydrodynamic or thermal timescale as
it adjusts to its new state.

The primary problem with this scenario is the timescales of the
brightness changes.  Since the thermal timescale for an early F star is $\sim 10^6$ yr, a
change of 15\% in only 100 years (or 3\% in 4 years) is about four orders of magnitude too
fast.  It is possible that a detailed hydrodynamical or other stellar structure simulation might reveal changes
on faster timescales, however, saving the hypothesis and suggesting an
origin of the dips.

This possibility finds some support in the consistency of the depth of
the sodium features, the measured reddening, and the
\citeauthor{Schlafly11} estimate of the reddening of 
Boyajian's Star.  Together, these argue that the total interstellar extinction
currently exhibited by Boyajian's Star is not especially higher than
expected for stars at this distance in this part of the {\it Kepler}
field. Since the star was apparently $\sim 17$\% brighter in 1890,
this implies that it was indeed overluminous then.

\section{Summary of Possibilities and Future Work}

Above, we have discussed several possible solutions to the problem of
Boyajian's Star.  Here, we list
them in rough order of our qualitative assessment of their
plausibility, along with a summary
of how they might explain the dips and the secular dimming and how future studies that could
help bolster them or rule them out.

{\it \S \ref{ISM} Small-scale ISM Structure --- plausible:}  Dense
regions of the ISM that vary on scales of au are known to exist and
cause phenomena similar to that seen in Boyajian's Star, though never
before noticed on these physical or column density scales.  This
possibility would find support if neighboring stars 
could be found to exhibit similar behavior to Boyajian's Star, or if
small-scale structure in the ISM in this direction could be found.
Similarly, this hypothesis would be strengthened if future dimmings
are accompanied by variations in reddening and absorption features
associated with interstellar dust.

{\it \S \ref{Bok} An Intervening Dark Cloud --- plausible:}  Bok
Globules exist, and there are clouds at higher Galactic latitude than
Boyajian's Star. Their relatively smooth density profiles would
naturally explain
the secular dimming, and if they have significant sub-au structure,
this would explain the dips.  This possibility would find support is
similar ways to the small-scale ISM structure possibility, but
additionally if the column of neutral or molecular gas were
significantly higher in the direction of Boyajian's Star.

{\it \S \ref{intervening} An Intervening Disk --- less plausible:} A
chance alignment with the large disk of a dark object --- such as a black
hole --- could explain the observations.  Complex ring structure in the annular
disk sculpted by bodies within it could explain the dips, while an
overall diffuse component would explain the secular dimming.  This
solution is similar to the case of 1SWASP J140747.93-394542.6 \citep[][an apparent
ringed proto-planet around a pre-main-sequence
star]{Mamajek12,Kenworthy2015}, although we find it more likely that
the disk would be in the field than in orbit around Boyajian's Star (\S \ref{OrbitingDisk}).
This hypothesis would find support if a pattern to the dips and
secular dimmings show symmetries consistent with a disk (as
\citeauthor{Kenworthy2015} found for J1407) or if more detailed studies of black
hole frequencies and disks support our rough calculation that such an
alignment is not terribly unlikely in the {\it Kepler} field.

{\it \S \ref{Megastructures} Artificial Structures
  --- spherical swarm not likely; other geometries' plausibility unclear:} The millimeter upper limits put tight constraints on any
circumstellar solution invoking a spherical cloud responsible
for the observed $\sim 17$\% secular dimming. A spherical swarm of artificial structures
is not quite excluded if they operate near the Carnot limit at
$T\sim 65$ K, but more sensitive FIR-millimeter work will be able to rule this
possibility out.  Other geometries would find support if future dips
and secular dimmings prove to be achromatic, or if the GAIA distance
indicates Boyajian's Star already suffers significant achromatic extinction.  If future dips are
accompanied by reddening or absorption features consistent with
ordinary astrophysical extinction, this possibility would be very unlikely.

{\it \S \ref{SS}: A solar system Cloud --- unclear:} A cold cloud of
material at large distance from the Sun, perhaps from a slow cometary
collision or a geyser from a large body, could cause obscuration but
be otherwise very difficult to detect.  The primary difficulty here is that
the plausibility of such a cloud existing at all is unclear.  If it
existed, it could explain the dips if it were clumpy and the secular
dimming if it had a more diffuse component.  Such a cloud might also
orbit around Boyajian's Star, where it could cause similar effects.
The hypothesis that such a cloud orbits the Sun could find support if
stars near Boyajian's Star could be found to suffer similar effects.
In either case, the absorption spectrum of the material (presumably
containing ices) might distinguish it from interstellar dust or other materials.

{\it \S \ref{Return} Post-merger Return to Normal --- unclear:} If
Boyajian's Star suffered a brightening, we may be seeing a return to
normal brightness, rather than a secular ``dimming.''  In this
scenario, the dips might be due to internal restructuring of the star
on a hydrodynamical timescale or leftover material from a merger,
while the secular ``dimming'' occurs on 
a thermal timescale for perhaps only the outer layers of the star.  It
is unclear what the mechanism could be, but it is possible that a
merger event with a close companion could result in such a
brightening.  While we do not find this hypothesis sufficiently
concrete to be classified as ``plausible'' or otherwise, we find the
it to be worth considering further.  On the theory side, this possibility would find
support if a brightening mechanism could be identified, and if the dimming
could be explained by that mechanism.  On the observational side, this
hypothesis would find support if future dimming events were not
accompanied by the reddening or absorption features expected from
dust, or if they were accompanied by changes in the effective
temperature or other properties of Boyajian's Star.  A
larger-than-expected GAIA distance would also indicate that the star
is overluminous.

{\it \S \ref{Circumstellar} Circumstellar Material, such as Cometary
  Swarms --- plausible for some of the dips, very unlikely for the
  secular dimming:} \citeauthor{WTF} discussed
the difficulties with most such hypotheses, and the comet swarm
hypothesis may explain some of the dips.  Invoking circumstellar
material to explain the secular dimming seems less fruitful,
especially given the millimeter flux upper limits.

{\it \S \ref{Timescales} Pulsations and Other Structural Variability
  --- not likely}: The variety of timescales of dimmings observed and
lack of mechanism for such pulsations make this possibility unlikely.
This possibility would find support if mechanisms to generate such
variability are found and if future dimmings are accompanied by
changes in stellar parameters but not reddening or absorption features.

{\it \S \ref{Spots} Polar Spots --- not likely:} Polar spots
are neither expected nor seen around F stars such as Boyajian's Star.
The variety of timescales of the dimmings also seems inconsistent with
spots.  While this possibility seems unlikely, we also find it
difficult to rule out entirely.  This possibility would find support
if spectra during future dips revealed the spots through, for
instance, Doppler imaging \citep[e.g.][]{VogtStar} or variations in
effective temperature or other spectral features.

{\it \S \ref{Instrumental}: Instrumental Effects --- very unlikely:}
We agree with \citeauthor{WTF} that instrumental effects are very unlikely
to be the cause of the dips and find \citet{Montet16} persuasive that
at least some secular dimming occurs, meaning it would be an unlikely
coincidence for the similar effects seen by \citeauthor{Schaefer16} to be instrumental.  Independent
confirmation of the long-term dimming seen by \citet{Montet16} would
further rule out this scenario.

\section{Conclusions}

In response to \citeauthor{Montet16}'s strong encouragement to
generate alternative hypotheses for the extraordinary light curve of
Boyajian's Star, we have examined new and existing data and attempted
to survey the landscape of potential solutions for plausibility.

We have shown that the timings of the deepest dips exhibited by
Boyajian's Star appear consistent with being randomly distributed in time, and
so potential explanations should not be constrained by any perceived
periodicities in the {\it Kepler} data. We argue that the star's
secular dimming combined with a lack of long-wavelength
excess in the star's SED strongly constrains scenarios involving
circumstellar material.  In particular, we find that no more than
0.2\% of Boyajian's Star's flux is being intercepted by the absorbing
material, despite what appears to be at least a 15\% decrease in
total flux toward Earth.  

We find that scenarios involving a spherical swarm of artificial structures absorbing the material are only just barely
consistent with the data if they involve non-dissipative work done at
the maximum (Carnot) efficiency at 65 K---other temperatures and
lower efficiencies are ruled out, although other swarm geometries are not. 

We have briefly surveyed a range of explanations that do not invoke
circumstellar material, and find two broad categories worthy of further
consideration: an intervening Bok globule or other ISM overdensity, and an intervening stellar
remnant with a large disk.  We have shown that the star's color excess,
absorption lines due to interstellar sodium, and predicted extinction
due to interstellar dust are all consistent with $A_V\sim 0.34$, or
about twice the total amount of secular dimming observed to date.

Less compelling, but difficult to rule out, are intrinsic variations
due to spots, a ``return to normal'' from a temporary brightening (due to, perhaps,
a stellar merger) and a cloud of material in the outer solar system.  We find instrumental effects,
other intrinsic variation in Boyajian's Star, and obscuration by a
disk around an orbital companion to Boyajian's Star very unlikely to
be responsible.

We have identified several additional lines of research that may help
explain Boyajian's Star's light curve, including {\it JWST}
MIR--FIR observations; optical broadband and spectroscopic
observations during future dips; a study of the ISM toward Boyajian's
Star; a hunt for similar variations in
stars near on the sky to Boyajian's Star; and careful consideration of
the Gaia parallax.

\acknowledgements

We thank Ben Montet and Josh Simon for sharing their manuscript with
us in advance of submission.  We thank Eugene Chiang, Adam Leroy, Carl
Heiles, Marshall Perrin, Kimberly Cartier, \added{Mark Conde, Robin
Ciardullo, Eric Feigelson, Steven Desch, Howard Bond,} and Fabienne Bastien for useful discussions
that contributed ideas to this paper.  \added{We thank the anonymous referee
for a prompt and constructive report that sharpened our reasoning and improved
the paper,} and a pseudonymous blog commenter for bringing the
\citet{Cameron71} reference to our attention.  We thank B.\ J.\ Fulton
and Andrew Howard for the Keck/HIRES spectrum of the sodium D
features.  We thank Davide Gandolfi for providing the NEXTGEN spectra
in Fig.~\ref{SED}.  We thank Josh Simon\added{, Ben Montet, B.\ J.\ Fulton},
Rebekah Dawson, and Tabetha 
Boyajian for careful readings of and detailed comments and
suggestions for our manuscript. We thank Jason Curtis for his analysis
of Boyajian's Star's sodium lines.   

\deleted{We thank the following for being the first, as we can best recall, to
bring the following ideas to our attention:}

\deleted{Mark Conde: that, in light of a thorough search for such periods,
the significance of the 24.2d period is weak.}

\deleted{Robin Ciardullo: That the dimming could be intrinsic}

\deleted{Eric Feigelson: That ``orphan'' high-latitude clouds exist}

\deleted{Steven Desch: That effluents from a close-in rocky planet might be
responsible for the {\it Kepler} dips}

\deleted{Howard Bond: That some PNe show similarly extraordinary light curves}

\replaced{Some}{Many} of the ideas and possible solutions in this work are not
original to us; \replaced{we regret not being able to credit all of them to their originators. In}{in} many
cases a particular solution has been suggested multiple times
independently in private \replaced{discussions}{communication}, at public talks, and/or \replaced{on}{in}
social media.

\added{This research was partially supported by Breakthrough Listen, part of
the Breakthrough Initiatives sponsored by the Breakthrough Prize
Foundation.\footnote{http://www.breakthroughinitiatives.org}} The Center for Exoplanets and Habitable Worlds is supported by the Pennsylvania State University, the Eberly College of Science, and the
 Pennsylvania Space Grant Consortium. The spectrum of Boyajian's Star presented herein was obtained at the
W.M. Keck Observatory, which is operated as a scientific partnership
among the California Institute of Technology, the University of
California and the National Aeronautics and Space Administration. The
Observatory was made possible by the generous financial support of the
W.M. Keck Foundation. 

The authors wish to recognize and acknowledge the very significant
cultural role and reverence that the summit of \replaced{Mauna Kea}{Maunakea} has always
had within the indigenous Hawaiian community.  We are most fortunate
to have the opportunity to conduct observations from this mountain. 

\bibliographystyle{aasjournal}



\end{document}